\magnification=1200
\hfuzz=3pt
\hsize=12.5cm
\hoffset=0.32cm
\baselineskip=18pt
\voffset=\baselineskip
\nopagenumbers
\font\title=cmssdc10 at 20pt
\font\smalltitle=cmssdc10 at 14pt

\newfam\msamfam
\font\tenmsam=msam10

\font\sevenmsam=msam7

\font\fivemsam=msam5
\textfont\msamfam=\tenmsam
\scriptfont\msamfam=\sevenmsam
\scriptscriptfont\msamfam=\fivemsam

\let\math=\mathchardef
\def\hexa #1{\ifcase #1 0\or 1\or 2\or 3\or 4\or 5\or 6\or 7\or 8\or 9\or A\or B\or C\or D\or E\or F\fi}

\math\gtrsim="3\hexa\msamfam26
\math\lesssim="3\hexa\msamfam2E

\let\text=\textstyle

\def\RE{\mathop{\Re e}\nolimits}
\def\IM{\mathop{\Im m}\nolimits}
\def\sectionstyle{\smalltitle}
\newskip\beforesectionskip
\newskip\aftersectionskip
\beforesectionskip=4mm plus 1mm minus 1mm
\aftersectionskip=2mm plus .2mm minus .2mm
\newcount\mysectioncounter
\def\resetsections{\mysectioncounter=0}
\resetsections
\newcount\myeqcounter
\def\mysection#1\par{\par\removelastskip\penalty -250
\vskip\beforesectionskip
\global\advance\mysectioncounter by 1\noindent
\myeqcounter=0{\sectionstyle\the\mysectioncounter.
#1}\par
\nobreak\vskip\aftersectionskip}
\def\myeqno{\global\advance\myeqcounter by 1\eqno{(\the\mysectioncounter.\the\myeqcounter)}}
\def\mydispeqno{\global\advance\myeqcounter by 1\hfill\llap{(\the\mysectioncounter
.\the\myeqcounter)}}
\newinsert\footins
\skip\footins=12pt plus 4pt minus 4pt
\count\footins=1000
\dimen\footins=8in

\def\footnote#1#2{#1\insert\footins{
\leftskip=0pt\rightskip=0pt
\interlinepenalty=1000
\baselineskip=10pt plus 1pt
\textindent{#1}#2\penalty-1000
\vskip 4pt plus 2pt minus 2pt}}
\headline={\hfil\tenrm\folio\hfil}
\footline={\hfil}
\baselineskip=12pt
\parindent= 20pt
\centerline{\smalltitle Out of equilibrium generalized Stokes-Einstein relation:} 
\medskip
\centerline{\smalltitle determination of the effective temperature}
\medskip
\centerline{\smalltitle of an aging medium}
\vskip 1cm
\centerline{No\"elle POTTIER}
\bigskip
\centerline{\sl F\'{e}d\'{e}ration de Recherche CNRS 2438 ``Mati\`{e}re et Syst\`{e}mes Complexes'',\/}
\centerline{\sl Universit\'{e} Paris 7,\/}
\bigskip
\centerline{\sl and\/}
\bigskip
\centerline{\sl Groupe de Physique des Solides, CNRS UMR 7588,\/} 
\centerline{\sl Universit\'{e}s Paris 6 et Paris 7,\/} 
\centerline{\sl 2, place Jussieu, 75251 Paris Cedex 05, France\/}
\vskip 1.5cm
\noindent
{\smalltitle Abstract}
\bigskip
\noindent
We analyze in details how the anomalous drift and diffusion properties of a particle evolving in an aging medium can be
interpreted in terms of an effective temperature of the medium. From an experimental point of view, independent measurements of
the mean-square displacement and of the mobility of a particle immersed in an aging medium such as a colloidal glass give
access to an out of equilibrium generalized Stokes-Einstein relation, from which the effective temperature of the medium can
eventually be deduced. We illustrate the procedure on a simple model with power-law behaviours. 
\bigskip
\parindent=0pt
{\bf PACS:} 

05.40.-a Fluctuation phenomena, random processes, noise and Brownian motion

02.50.Ey Stochastic processes
\vskip 1cm
{\bf KEYWORDS}: Fluctuation phenomena\kern0.2em; Random processes\kern0.2em; Noise and Brownian motion\kern 0.2em; Stochastic
processes 
\bigskip
\hrule width15truecm
\bigskip
{\parskip=0pt
{\sl Corresponding author\/:
\smallskip
No\"elle POTTIER
 
Fax number\/: + 33 1 46 33 94 01 

E-mail\/: npottier@ccr.jussieu.fr\/}}
\vfill
\break
\parskip=0pt
\parindent=20pt
\baselineskip=16pt
\mysection{Introduction}

As well-known, the dynamical properties of systems which are not far from an equilibrium state can successfully be analyzed in
terms of the fluctuation-dissipa\-tion theorems (FDTs) [1,2]. As recalled for instance in [3], the meaning of these theorems is
as follows. The fluctuation-dissipation theorem of the ``first kind'' (or first FDT) expresses a necessary condition for a
thermometer in contact solely with the system to register the temperature of the bath. The fluctuation-dissipation theorem of
the ``second kind'' (or second FDT) expresses the fact that the bath itself is in equilibrium.

None of these theorems is valid out of equilibrium. This happens in particular when the bath is constituted
by an aging medium with properties evolving with its age ({\sl i.e.\/} the time $t_w$ elapsed since the instant of its
preparation). In such a case, provided that departures from equilibrium remain weak, it has been proposed to associate to
the FDT violation an age and frequency-dependent effective temperature $T_{\rm eff.}(\omega,t_w)$[4,5]. More precisely, the
effective temperature has been introduced as a quantity modifying the relation between the dissipative part of a
given generalized susceptibility and the associated correlation function. A few experimental measurements of the effective
temperature of real out of equilibrium physical systems have subsequently been performed [6,7,8,9].

In the present paper, we focus the interest on the recent experiment as described in [9], in which the effective temperature of
an aging colloidal glass is determined by studying the mean square displacement and the frequency-dependent mobility of an
immersed probe particle. If the particle would evolve in a bath at equilibrium, the diffusion exponent characterizing its mean
square displacement $\Delta x^2(t)$ would only depend on the exponent characterizing the real part $\RE\mu(\omega)$ of its
mobility [10]. For an out of equilibrium bath as characterized by an effective temperature $T_{\rm
eff.}(\omega)$, we showed in [11] that, if the latter can be modelized by an inverse power-law of $\omega$ at small
$\omega$, the diffusion exponent depends on both exponents associated with $\RE\mu(\omega)$ and $T_{\rm eff.}(\omega)$. Then,
in turn, independent measurements of the exponents characterizing $\RE\mu(\omega)$ and $\Delta x^2(t)$ allow for the
determination of the exponent associated with the effective temperature. 

Here we present a more thorough analysis of this problem. We show how the full expression of the effective temperature can
be deduced from the measurements of the particle mean square displacement and mobility. This general analysis does not rely on
any specific hypothesis on the behaviours of $\Delta x^2(t)$ and $\mu(\omega)$. We show that, in the aging medium, the
particle mean-square displacement and mobility are linked together {\sl via\/} an out of equilibrium generalized
Stokes-Einstein relation, which itself derives from a modified Kubo formula for $\mu(\omega)$. The question of the
determination of $T_{\rm eff.}(\omega)$ is not straightforward since the modified Einstein relation defining the effective
temperature solely involves $\RE\mu(\omega)$. 

The paper is organized as follows. In Sections 2 and 3, we recall the formulation of the equilibrium FDTs. We put the emphasis
on the fact that, for each one of the two FDTs, two equivalent formulations can be given. One formulation concerns the
dissipative part of a given generalized susceptibility, while the other one concerns the corresponding Kubo formula. In
Section 4, we turn to out of equilibrium situations. We show that the effective temperature $T_{\rm eff.}(\omega)$, as defined
{\sl via\/} the modified Einstein and Nyquist relations, is not identical to the quantities ${\cal T}^{(1)}_{\rm
eff.}(\omega)$ or ${\cal T}^{(2)}_{\rm eff.}(\omega)$ involved in the corresponding modified Kubo formulas. We establish the
relation linking $T_{\rm eff.}(\omega)$ and ${\cal T}^{(1)}_{\rm eff.}(\omega)$ (and also the one linking $T_{\rm
eff.}(\omega)$ and ${\cal T}^{(2)}_{\rm eff.}(\omega)$). In Section 5, we show how this relation allows to deduce the
effective temperature from the out of equilibrium generalized Stokes-Einstein relation linking the Laplace transform of
the mean square displacement and the $z$-dependent mobility ($z$ denotes the Laplace variable). In Section
6, we illustrate the whole procedure by considering again the model with power-law behaviours previously studied in [11].
Finally, Section 7 contains our conclusions.

\mysection {Diffusion in a stationary medium}

The motion of a diffusing particle of mass $m$ evolving in a stationary medium is usually described by the generalized
Langevin equation [1,2],
$$m\,{dv\over dt}=-m\int_{-\infty}^\infty\tilde\gamma(t-t')\,v(t')\,dt'+F(t),\qquad v={dx\over dt},\myeqno$$
in which $F(t)$ is the Langevin random force acting on the particle and $\tilde\gamma(t)$ a retarded friction kernel
accounting for the viscoelastic properties of the medium. In Eq. (2.1), it is assumed that the diffusing particle and its
stationary environment have been put in contact in an infinitely remote past, as pictured by the lower integration
bound $-\infty$ in the retarded friction term. Both $F(t)$ and the solution $v(t)$ of Eq. (2.1) can be viewed as stationary
random processes. Their spectral densities are linked by
$$C_{vv}(\omega)=|\mu(\omega)|^2\,C_{FF}(\omega),\myeqno$$
where
$$\mu(\omega)={1\over m[\gamma(\omega)-i\omega]}\myeqno$$
denotes the frequency-dependent particle mobility. In Eq. (2.3), the friction coefficient $\gamma(\omega)$ is the Fourier
transform of the retarded kernel $\tilde\gamma(t)$, as defined by
$\gamma(\omega)=\int_{-\infty}^\infty\tilde\gamma(t)\,e^{i\omega t}\,dt$. 
\bigskip
\mysection{Formulations of the equilibrium fluctuation-dissipation theorems}

When the stationary medium surrounding the diffusing particle is in thermal equilibrium at temperature $T$,
specific relations exist, which allow to express the mobility in terms of the velocity correlation function, and the friction
coefficient in terms of the random force correlation function. These two relations respectively constitute the first and the
second FDTs [1,2]. 

Interestingly, each one of the two FDTs can be formulated in two equivalent ways, depending on whether
one is primarily interested in writing a Kubo formula for a generalized susceptibility $\chi(\omega)$ (in the present case
$\mu(\omega)$ or $\gamma(\omega)$), or an expression for its dissipative part (namely the Einstein relation for
$\RE\mu(\omega)$ or the Nyquist formula for $\RE\gamma(\omega)$). This feature is far from being academic: indeed the
equivalence between these two types of formulations, which holds at equilibrium, has to be carefully reconsidered out of
equilibrium, when one attempts to extend the FDTs with the help of a frequency-dependent effective temperature [4,5].

Before studying out of equilibrium situations, let us briefly summarize the two formulations of each of the
equilibrium theorems.
\bigskip
{\bf 3.1. Formulations of the first fluctuation-dissipation theorem}

The linear response theory, applied to the particle
velocity considered as a dynamical variable of the isolated particle-plus-bath system, allows to express the mobility
$\mu(\omega)=\chi_{vx}(\omega)$ in terms of the equilibrium velocity correlation function: 
$$\mu(\omega)={1\over kT}\int_0^\infty\bigl\langle v(t)v\bigr\rangle\,e^{i\omega
t}\,dt.\myeqno$$ 
The Kubo formula (3.1) for the mobility constitutes a formulation of the first FDT [1,2].

Accordingly, the velocity spectral density is related to the real part of the mobility:
$$C_{vv}(\omega)=\int_{-\infty}^\infty\bigl\langle v(t)v\bigr\rangle\,e^{i\omega t}\,dt=kT\,2\RE\mu(\omega).\myeqno$$
Introducing the frequency-dependent diffusion coefficient $D(\omega)$ as defined by
$$D(\omega)=\int_0^\infty\bigl\langle v(t)v\bigr\rangle\,\cos\omega t\,dt,\myeqno$$
one can rewrite Eq. (3.2) as the well-known Einstein relation:
$${D(\omega)\over\RE\mu(\omega)}=kT.\myeqno$$

The Einstein relation (3.4) or the expression (3.2) of the dissipative part $\RE\mu(\omega)$ of the mobility constitute
another formulation of the first FDT. Indeed they contain the same information as the Kubo formula (3.1), since $\mu(\omega)$
can be deduced from $\RE\mu(\omega)$ with the help of the usual Kramers-Kronig relations valid for real $\omega$ [1,2]. Eq.
(3.1) on the one hand, and Eqs. (3.2) or (3.4) on the other hand, are thus fully equivalent, and they all involve
the thermodynamic bath temperature $T$. Note however that, while $\mu(\omega)$ as given by Eq. (3.1) can be extended into
an analytic function in the upper complex half plane, the same property does not hold for $D(\omega)$.
\bigskip
{\bf 3.2. Formulations of the second fluctuation-dissipation theorem}

From the first FDT (3.2) and the expression (2.3) of the mobility $\mu(\omega)$, one gets:
$$C_{vv}(\omega)={kT\over m}\,{2\RE\gamma(\omega)\over|\gamma(\omega)-i\omega|^2}.\myeqno$$
One then deduces from Eq. (2.2) the well-known Nyquist formula yielding the noise spectral density in terms of the dissipative
part $\RE\gamma(\omega)$ of the friction coefficient:
$$C_{FF}(\omega)=mkT\,2\RE\gamma(\omega).\myeqno$$

Correspondingly, one can write a Kubo formula relating the friction coefficient $\gamma(\omega)$ to the random force
correlation function:
$$\gamma(\omega)={1\over mkT}\int_0^\infty \bigl\langle F(t)F\bigr\rangle\,e^{i\omega t}\,dt.\myeqno$$
Eq. (3.6) and Eq. (3.7) are two equivalent formulations of the second FDT, which both involve the
bath temperature $T$ [1,2]. 

\mysection{Effective temperature in an out of equilibrium medium: the link with the Kubo formulas for the generalized
susceptibilities}

Let us now consider a particle diffusing in an out of equilibrium environment, for which no well defined thermodynamical
temperature does exist. Since, in an out of equilibrium regime, even stationary, the FDTs are not
satisfied, one can try to rewrite them in a modified way, and thus to extend the linear response theory, with the help of a
(frequency-dependent) effective temperature. 

Such a quantity, denoted as $T_{\rm eff.}(\omega)$, and parametrized by the age of the system, has been defined, for real
$\omega$, {\sl via\/} an extension of both the Einstein relation and the Nyquist formula. It has been argued in [4,5] that the
effective temperature defined in this way plays in out of equilibrium systems the same role as the
thermodynamic temperature in systems at equilibrium (namely, the effective temperature controls the direction of heat flow and
acts as a criterion for thermalization).

The effective temperature can in principle be deduced from independent measurements, for instance, of
$\RE\mu(\omega)$ and $D(\omega)$ (or of $\RE\gamma(\omega)$ and $C_{FF}(\omega)$). However, experimentally it may be
preferable to make use of the modified Kubo formulas for the corresponding generalized susceptibilities.  The Kubo formula for
$\mu(\omega)$ (and also the one for
$\gamma(\omega)$) cannot be extended to an out of equilibrium situation just replacing $T$ by $T_{\rm eff.}(\omega)$ in Eq.
(3.1) (and in Eq. (3.7)) [11]. In the following, we will show in details how the Kubo formulas have then to be rewritten. 
\bigskip
{\bf 4.1. Definition of the effective temperature $T_{\rm eff.}(\omega)$}

To begin with, following [4,5], we write the relation between the dissipative part of the mobility and the velocity
spectral density in an out of equilibrium medium as:
$$C_{vv}(\omega)=kT_{\rm eff.}(\omega)\,2\RE\mu(\omega).\myeqno$$
This amounts to assume a modified Einstein relation:
$${D(\omega)\over\RE\mu(\omega)}=kT_{\rm eff.}(\omega).\myeqno$$
Eq. (4.1) (or Eq. (4.2)) defines a frequency-dependent effective temperature $T_{\rm eff.}(\omega)$. 

Interestingly, the same effective temperature $T_{\rm eff.}(\omega)$ can consistently be used in the modified Nyquist formula
linking the noise spectral density
$C_{FF}(\omega)=\int_{-\infty}^\infty\langle F(t)F\rangle e^{i\omega t}dt$ and the dissipative part $\RE\gamma(\omega)$ of the
friction coefficient. One has:
$$C_{FF}(\omega)=mkT_{\rm eff.}(\omega)\,2\RE\gamma(\omega).\myeqno$$

Let us note for further purpose that $T_{\rm eff.}(\omega)$ is only defined for $\omega$ real. Eqs. (4.1)-(4.3)
cannot be extended to complex values of $\omega$. 
\bigskip
{\bf 4.2. The modified Kubo formula for the mobility}

Out of equilibrium, it is not possible to rewrite the Kubo formula for $\mu(\omega)$ under a form similar to Eq.
(3.1) with
$T_{\rm eff.}(\omega)$ in place of $T$:
$$\mu(\omega)\neq{1\over kT_{\rm eff.}(\omega)}\int_0^\infty\bigl\langle v(t)v\bigr\rangle\,e^{i\omega
t}\,dt.\myeqno$$  
Let us instead write a modified Kubo formula as: 
$$\mu(\omega)={1\over k{\cal T}_{\rm eff.}^{(1)}(\omega)}\int_0^\infty\bigl\langle
v(t)v\bigr\rangle\,e^{i\omega t}\,dt.\myeqno$$ Eq. (4.5) defines a complex frequency-dependent function ${\cal
T}_{\rm eff.}^{(1)}(\omega)$, in terms of which the velocity spectral density
$C_{vv}(\omega)=\int_{-\infty}^\infty\langle v(t)v\rangle e^{i\omega t}dt$ writes:
$$C_{vv}(\omega)=2\RE\bigl[\mu(\omega)k{\cal T}_{\rm eff.}^{(1)}(\omega)\bigr].\myeqno$$
According to Eq. (4.5), the quantity $\mu(\omega)k{\cal T}_{\rm
eff.}^{(1)}(\omega)=\int_0^\infty\langle v(t)v\rangle e^{i\omega t}\,dt$ can be extended into an analytic function in the
upper complex half plane. Since this analyticity property holds for $\mu(\omega)$, it also holds for ${\cal T}_{\rm
eff.}^{(1)}(\omega)$. 

For real $\omega$, one has the identity:
$$T_{\rm eff.}(\omega)\,2\RE\mu(\omega)=2\RE\bigl[\mu(\omega){\cal T}_{\rm eff.}^{(1)}(\omega)\bigr].\myeqno$$
Eq. (4.7) allows to derive the function ${\cal T}_{\rm
eff.}^{(1)}(\omega)$ from a given effective temperature $T_{\rm eff.}(\omega)$. Conversely, it provides the link
allowing for deducing $T_{\rm eff.}(\omega)$ from the modified Kubo formula for $\mu(\omega)$.

Thus, the velocity correlation function, which is the inverse Fourier transform of $C_{vv}(\omega)$, can be expressed in
two equivalent forms, either, in terms of the effective temperature, as
$$\bigl\langle v(t)v(t')\bigr\rangle=\int_{-\infty}^\infty{d\omega\over
2\pi}\,e^{-i\omega(t-t')}\,kT_{\rm eff.}(\omega)\,
2\RE\mu(\omega),\myeqno$$
or, in terms of the function ${\cal T}_{\rm eff.}^{(1)}(\omega)$, as:
$$\bigl\langle v(t)v(t')\bigr\rangle=\int_{-\infty}^\infty{d\omega\over
2\pi}\,e^{-i\omega(t-t')}\,2\,\RE\bigl[\mu(\omega)k{\cal T}_{\rm eff.}^{(1)}(\omega)\bigr].\myeqno$$
Let us note for further purpose that, since $C_{vv}(\omega)$ is an even function of $\omega$, Eq. (4.9)
can be rewritten as a cosine Fourier integral,
$$\bigl\langle v(t)v(t')\bigr\rangle=\int_{-\infty}^\infty{d\omega\over
2\pi}\,\cos\omega(t-t')\,2\,\RE\bigl[\mu(\omega)k{\cal T}_{\rm eff.}^{(1)}(\omega)\bigr],\myeqno$$
that is\footnote{$^1$}{Since $\RE[\mu(\omega)k{\cal T}_{\rm eff.}^{(1)}(\omega)]$ is even,
$\IM[\mu(\omega)k{\cal T}_{\rm eff.}^{(1)}(\omega)]$ is odd, as it can be deduced from the Kramers-Kronig
relations linking these two quantities. For this reason, Eqs. (4.10) and (4.11) are equivalent.}:
$$\bigl\langle v(t)v(t')\bigr\rangle=\int_{-\infty}^\infty{d\omega\over
2\pi}\,\cos\omega(t-t')\,2\,\mu(\omega)k{\cal T}_{\rm eff.}^{(1)}(\omega).\myeqno$$
\bigskip
{\bf 4.3. The modified Kubo formula for the friction coefficient}

Similarly, out of equilibrium, it is not possible to rewrite the Kubo formula for $\gamma(\omega)$ under a form similar to Eq.
(3.7) with $T_{\rm eff.}(\omega)$ in place of $T$:
$$\gamma(\omega)\neq{1\over mkT_{\rm eff.}(\omega)}\int_0^\infty\bigl\langle F(t)F\bigr\rangle\,e^{i\omega t}\,dt.\myeqno$$
Let us instead write a modified Kubo formula as:
$$\gamma(\omega)={1\over mk{\cal T}_{\rm eff.}^{(2)}(\omega)}\int_0^\infty\bigl\langle F(t)F(t')\bigr\rangle\,e^{i\omega
t}\,dt.\myeqno$$
One has:
$$C_{FF}(\omega)=2m\RE\bigl[\gamma(\omega)k{\cal T}_{\rm eff.}^{(2)}(\omega)\bigr].\myeqno$$
According to Eq. (4.13), the quantity $m\gamma(\omega)k{\cal T}_{\rm
eff.}^{(2)}(\omega)=\int_0^\infty\langle F(t)F\rangle\,e^{i\omega t}\,dt$ can be extended into an analytic function in the
upper complex half plane. The same property holds for ${\cal T}_{\rm eff.}^{(2)}(\omega)$.

For real $\omega$, one has the identity:
$$T_{\rm eff.}(\omega)\,2\RE\gamma(\omega)=2\RE\bigl[\gamma(\omega){\cal T}_{\rm eff.}^{(2)}(\omega)\bigr].\myeqno$$
Eq. (4.15) allows to derive the function ${\cal T}_{\rm
eff.}^{(2)}(\omega)$ from a given effective temperature $T_{\rm eff.}(\omega)$, and conversely.

Note that, although the same effective temperature $T_{\rm eff.}(\omega)$ can be consistently used in both the modified
Einstein relation (4.2) and the modified Nyquist formula (4.3), the quantities ${\cal T}_{\rm eff.}^{(1)}(\omega)$ and ${\cal
T}_{\rm eff.}^{(2)}(\omega)$ involved in the corresponding Kubo formulas ({\sl i.e.\/} Eq. (4.5) for $\mu(\omega)$ and Eq.
(4.13) for $\gamma(\omega)$) are not identical. 

\mysection{The out of equilibrium generalized Stokes-Einstein relation and the determination of the effective temperature}

Let us consider a particle evolving in an out of equilibrium environment (for instance, an aging colloidal glass). In a
quasi-stationary regime, as pictured by the condition $\omega t_w\gg 1$, it exists two  well separated time scales,
respectively, characterizing the times pertinent for the measuring process (of order $\omega^{-1}$), and the aging time
$t_w$. At a given $t_w$, the particle velocity can then be assumed to obey the generalized Langevin equation (2.1). 

Given the real part $\RE\gamma(\omega)$ of the friction coefficient and the effective temperature
$T_{\rm eff.}(\omega)$, the spectral density $C_{FF}(\omega)$ can be derived from the modified Nyquist formula  (4.3). Then,
Eq. (2.1) can be solved ({\sl i.e.\/} in particular $D(\omega)$ can be obtained).  Conversely, independent measurements of
the particle mean-square displacement and mobility in the aging medium allow for the determination of its effective
temperature. The aim of the present Section is to present a simple and convenient way of deducing $T_{\rm eff.}(\omega)$ from
the experimental results.

Experimentally, the effective temperature of a colloidal glass can be determined by studying the anomalous drift and diffusion
properties of an immersed probe particle. More precisely, one measures, at the same age of the medium, the
particle mean-square displacement as a function of time on the one hand, and its frequency-dependent mobility on the other
hand. This program has recently been achieved for a micrometric bead immersed in a glassy colloidal suspension of Laponite. As
a result, both $\Delta x^2(t)$ and $\mu(\omega)$ are found to display power-law behaviours [9].
\bigskip
{\bf 5.1. Asymptotic analysis determination of the effective temperature in a model with power-law behaviours} 

In a previous paper [11], we proposed a preliminary study of this problem. We considered an aging medium characterized, at
small frequencies, by a friction coefficient proportional to $|\omega|^{\delta-1}$ (with $0<\delta<2$). The diffusion
properties were characterized by an exponent $\nu$ (with $0<\nu<2$ since the motion must not be kinematical). In other words,
at large times the particle mean-square displacement was assumed to behave like $\Delta x^2(t)\sim t^\nu$ (the case $0<\nu<1$
corresponds to a sub-diffusive behaviour, while the case $1<\nu<2$ corresponds to a super-diffusive one).

At equilibrium, one has $\nu=\delta$ [10]. Out of equilibrium, there is no {\sl a priori\/} specific relation
between $\nu$ and $\delta$. Instead we showed, by applying asymptotic Fourier analysis to the expression (4.8) of $\langle
v(t)v(t')\rangle$ in terms of $T_{\rm eff.}(\omega)$, that the aging medium can be characterized by an effective temperature
behaving like $|\omega|^\alpha$ at low frequencies, with: 
$$\alpha=\delta-\nu.\myeqno$$
This asymptotic analysis directly relies on the modified Einstein relation (4.2). It does not necessitate the use of the
modified Kubo formula (4.5) [11]. 
\bigskip
{\bf 5.2. The out of equilibrium generalized Stokes-Einstein relation}

In the present paper, we want to go one step further and to show how the modified Kubo formula (4.5) for $\mu(\omega)$
leads to a relation between the (Laplace transformed) mean-square displacement and the $z$-dependent mobility ($z$ denotes
the Laplace variable). This out of equilibrium generalized Stokes-Einstein relation makes an explicit use of the function
${\cal T}_{\rm eff.}^{(1)}(\omega)$ involved in the modified Kubo formula (4.5), a quantity which is not identical to the
effective temperature $T_{\rm eff.}(\omega)$, but from which $T_{\rm eff.}(\omega)$ can be deduced using the identity (4.7).
Interestingly, this way of obtaining the effective temperature is fully  general ({\sl i.e.\/} it is not restricted to large
times and small frequencies). It is therefore well adapted to the analysis of the experimental results [9]. 

The mean-square displacement of the diffusing particle,
$$\Delta x^2(t)=\bigl\langle[x(t)-x(t=0)]^2\bigr\rangle,\qquad t>0,\myeqno$$
 can be deduced from the velocity correlation function {\sl via\/} a double integration over time:
$$\Delta x^2(t)=2\int_0^t dt_1\int_0^{t_1}dt_2\,\bigl\langle v(t_1)v(t_2)\bigr\rangle.\myeqno$$

Introducing the Laplace transformed quantities $\hat v(z)=\int_0^\infty v(t)\,e^{-zt}\,dt$ and
$\widehat{\Delta x^2}(z)=\int_0^\infty\Delta x^2(t)\,e^{-zt}\,dt$, one gets, by Laplace transforming Eq. (5.2):
$$z^2\,\widehat{\Delta x^2}(z)=2\,\bigl\langle\hat v(z)v(t=0)\bigr\rangle.\myeqno$$
By Laplace transforming the expression of $\langle v(t)v(t=0)\rangle$ as a cosine Fourier transform of $C_{vv}(\omega)$, one
obtains:
$$\bigl\langle\hat v(z)v(t=0)\bigr\rangle=\int_{-\infty}^\infty{d\omega\over 2\pi}\,{z\over
z^2+\omega^2}\,C_{vv}(\omega).\myeqno$$
Introducing in Eq. (5.5) the expression (4.6) of $C_{vv}(\omega)$ (and using the fact that
$\IM[\mu(\omega)k{\cal T}_{\rm eff.}^{(1)}(\omega)]$ is an odd function of $\omega$), one gets:
$$z^2\,\widehat{\Delta x^2}(z)=2\int_{-\infty}^\infty{d\omega\over 2\pi}\,{z\over z^2+\omega^2}\,2\,\mu(\omega)k{\cal T}_{\rm
eff.}^{(1)}(\omega).\myeqno$$ 
Eq. (5.6) yields, by standard contour integration:
$$z^2\,\widehat{\Delta x^2}(z)=2\hat\mu(z)k\,\hat{\cal T}_{\rm eff.}^{(1)}(z).\myeqno$$
In Eq. (5.7), one has introduced the $z$-dependent quantities:
$$\hat\mu(z)=\mu(\omega=iz),\qquad\hat{\cal T}_{\rm eff.}^{(1)}(z)={\cal T}_{\rm eff.}^{(1)}(\omega=iz).\myeqno$$
Eq. (5.7) is the extension to an out of equilibrium medium of the generalized Stokes-Einstein relation
$z^2\,\widehat{\Delta x^2}(z)=2\hat\mu(z)kT$, valid for a medium in equilibrium [12]. 
Let us emphasize that the function ${\cal T}_{\rm eff.}^{(1)}(\omega)$ is not the effective temperature involved in both the
modified Einstein relation (4.2) and Nyquist formula (4.3). Instead it is the quantity which appears in the modified Kubo
formula for the mobility (Eq. (4.5)). Let us add that one could also have obtained Eq. (5.7) from the Kubo formula (4.5) by
setting $z=-i\omega$ in this latter relation (which is allowed since the function $\mu(\omega){\cal T}_{\rm
eff.}^{(1)}(\omega)$ is analytic in the upper complex half plane):
$$\langle\hat v(z)v(t=0)\rangle=\hat\mu(z)k\hat{\cal T}_{\rm eff.}^{(1)}(z).\myeqno$$
Eq. (5.7) then directly follows (using Eq. (5.4)).
\bigskip
{\bf 5.3. The effective temperature}

As displayed by the out of equilibrium generalized Stokes-Einstein relation (5.7), independent measurements of the particle
mean-square displacement and frequency-dependent mobility in an aging medium give access, once $\widehat{\Delta x^2}(z)$ and
$\hat\mu(z)=\mu(\omega=iz)$ are determined, to
$\hat{\cal T}_{\rm eff.}^{(1)}(z)$, and to ${\cal T}_{\rm eff.}^{(1)}(\omega)=\hat{\cal T}_{\rm eff.}^{(1)}(z=-i\omega)$.
Then, the identity (4.7) yields the effective temperature:
$$T_{\rm eff.}(\omega)={\RE\bigl[\mu(\omega){\cal T}_{\rm eff.}^{(1)}(\omega)\bigr]\over \RE\mu(\omega)}.\myeqno$$
Formula (5.10), together with the out of equilibrium generalized Stokes-Einstein relation
(5.7), is the central result of the present Section.

The above described method has recently been used in [9] to obtain the effective temperature of a glassy colloidal
suspension of Laponite through an immersed micrometric bead diffusion and mobility measurements.

\mysection{Determination of the effective temperature in a model with power-law behaviours}

To illustrate the whole procedure, let us consider again the previously described model, in which both $\Delta x^2(t)$ and
$\mu(\omega)$ are assumed to display power-law behaviours. 
\bigskip
{\bf 6.1. The friction coefficient and the mobility}

We take for $\RE\gamma(\omega)$ (for $\omega$ real) a function behaving like a power-law of exponent $\delta-1$
with $0<\delta<2$ [10,11]:
$$\RE\gamma(\omega)=\gamma_\delta\,\Bigl({|\omega|\over\tilde\omega}\Bigr)^{\delta-1},\qquad|\omega|\ll\omega_c.\myeqno$$ 
In Eq. (6.1), $\omega_c$ denotes a cut-off frequency typical of the environment, and $\tilde\omega\ll\omega_c$ is a reference
frequency allowing for the coupling constant $\eta_\delta=m\gamma_\delta$ to have the dimension of a viscosity for any
$\delta$. Introducing the $\delta$-dependent characteristic frequency $\omega_\delta$ as defined by
$$\omega_\delta^{2-\delta}=\gamma_\delta\,{1\over\tilde\omega^{\delta-1}}\,{1\over\sin{\delta\pi\over 2}},\myeqno$$
one has:
$$\RE\gamma(\omega)=\omega_\delta\,\Bigl({|\omega|\over\omega_\delta}\Bigr)^{\delta-1}\,\sin{\delta\pi\over
2},\qquad|\omega|\ll\omega_c.\myeqno$$
The function $\gamma(\omega)$ can be defined for any complex $\omega$ (except for a cut on the real negative axis).
It is given by:
$$\gamma(\omega)=\omega_\delta\,\Bigl({-i\omega\over\omega_\delta}\Bigr)^{\delta-1},\qquad|\omega|\ll\omega_c,\qquad
-\pi<\arg\omega\leq\pi.\myeqno$$

The mobility $\mu(\omega)$ is related to $\gamma(\omega)$ by Eq. (2.3). In the frequency range
$|\omega|\ll\omega_\delta$, inertia can be neglected. Then, we may use the overdamped mobility:
$$\mu(\omega)\simeq{1\over m\gamma(\omega)},\qquad|\omega|\ll\omega_\delta.\myeqno$$
One has, for $\omega$ real:
$$\RE\mu(\omega)\simeq{1\over
m\omega_\delta}\,\Bigl({|\omega|\over\omega_\delta}\Bigr)^{1-\delta}\,\sin{\delta\pi\over
2},\qquad|\omega|\ll\omega_\delta.\myeqno$$ 
For $\omega$ complex (except for a cut on the real negative axis), one has:
$$\mu(\omega)\simeq{1\over m\omega_\delta}\,\Bigl({-i\omega\over\omega_\delta}\Bigr)^{1-\delta},\qquad
-\pi<\arg\omega\leq\pi,\qquad|\omega|\ll\omega_\delta.\myeqno$$
From Eq. (6.7), one deduces the function $\hat\mu(z)$:
$$\hat{\mu}(z)={1\over m\omega_\delta}\,\Bigl({z\over\omega_\delta}\Bigr)^{1-\delta},\qquad|z|\ll\omega_\delta.\myeqno$$ 
\bigskip
{\bf 6.2. The mean-square displacement}

Let us set
$$\Delta x^2(t)\sim 2\,{kT\over m\omega_\nu^2}\,{(\omega_\nu t)^\nu\over\Gamma(\nu+1)},\qquad\omega_\nu t\gg 1,\myeqno$$
where $\omega_\nu^{-1}$ denotes some characteristic time after which the behaviour (6.9) is well settled.
The behaviour of the mean-square displacement is a (possibly anomalous) diffusive one, as characterized by the
exponent $\nu$ ($0<\nu<2$). Correspondingly, one has:
$$\widehat{\Delta x^2}(z)=2\,{kT\over m\omega_\nu^2}\,\omega_\nu^{\nu}\,z^{-\nu-1},\qquad|z|\ll\omega_\nu.\myeqno$$
\bigskip
{\bf 6.3. Determination of the effective temperature}

From the out of equilibrium generalized Stokes-Einstein relation (5.7), together with the expressions of $\hat\mu(z)$
(Eq. (6.8)) and of $\widehat{\Delta x^2}(z)$ (Eq. (6.10)), one deduces
$$\hat{\cal T}_{\rm
eff.}^{(1)}(z)=T\,{\omega_\delta^{2-\delta}\over\omega_\nu^{2-\nu}}\,z^{\delta-\nu},\qquad
|z|\ll\omega_\delta,\omega_\nu,\myeqno$$  
and:
$${\cal T}_{\rm
eff.}^{(1)}(\omega)=T\,{\omega_\delta^{2-\delta}\over\omega_\nu^{2-\nu}}\,(-i\omega)^{\delta-\nu},
\qquad|\omega|\ll\omega_\delta,\omega_\nu.\myeqno$$
Then, using Eq. (5.10) and the expressions (6.7) of $\mu(\omega)$ and (6.12) of ${\cal T}_{\rm
eff.}^{(1)}(\omega)$, one obtains the effective temperature of the aging medium:
$$T_{\rm eff.}(\omega)=T\,{\omega_\delta^{2-\delta}\over\omega_\nu^{2-\nu}}\,{\sin{\nu\pi\over 2}\over\sin{\delta\pi\over
2}}\,|\omega|^{\delta-\nu},\qquad|\omega|\ll\omega_\delta,\omega_\nu.\myeqno$$
Setting $\alpha=\delta-\nu$, and introducing the characteristic frequency $\omega_0$ as defined by
$$\omega_0^\alpha={\omega_\delta^{\delta-2}\over\omega_\nu^{\nu-2}}\,{\sin{\delta\pi\over 2}\over\sin{\nu\pi\over 2}},\myeqno$$
one can write:
$$T_{\rm eff.}(\omega)=T\,\Bigl({|\omega|\over\omega_0}\Bigr)^\alpha,\qquad|\omega|\ll\omega_0.\myeqno$$
The effective temperature behaves like a power-law of $\omega$ with an exponent $\alpha=\delta-\nu$.

We recover, as expected, the results previously obtained in [11] through the asymptotic Fourier analysis of the velocity
correlation function. The present method avoids completely the use of correlation functions, and makes use only of
one-time quantities ({\sl via\/} their Laplace transforms). It is particularly well suited to the interpretation of numerical
data.  

\mysection{Conclusion}

The main new results of this paper are the out of equilibrium generalized Stokes-Einstein relation between $\widehat{\Delta
x^2}(z)$ and $\hat{\mu}(z)$, together with the formula linking $T_{\rm eff.}(\omega)$ and the
quantity, denoted as ${\cal T}^{(1)}(\omega)$, involved in the Stokes-Einstein relation. One thus has at hand an efficient way
of deducing the effective temperature from the experimental results [9].

The above described procedure represents an actual improvement over the one previously presented in [11], which relied upon an
asymptotic Fourier analysis of the relation between the velocity correlation function (as deduced from the mean square
displacement) and the mobility. Indeed the latter method was restricted to large times and small frequencies, while the one
presented here makes it possible to take into account the behaviour of the mean square displacement and mobility in the full
range of experimentally accessible values.
\vfill
\break
\noindent
{\smalltitle Acknowledgements}

We wish to thank B. Abou and F. Gallet for helpful discussions on this question.
\bigskip
\parindent=0pt
{\smalltitle References}
\bigskip
\baselineskip=12pt
\frenchspacing
 
1. R. Kubo, Rep. Prog. Phys. {\bf 29}, 255 (1966).

2. R. Kubo, M. Toda and N. Hashitsume, {\it Statistical physics
\uppercase\expandafter{\romannumeral 2} : nonequilibrium statistical mechanics\/}, Second edition, Springer-Verlag, Berlin,
1991.

3. L.F. Cugliandolo, D.S. Dean and J. Kurchan, Phys. Rev. Lett. {\bf 79}, 2168 (1997).

4. L.F. Cugliandolo, J. Kurchan and L. Peliti, Phys. Rev. E {\bf 55}, 3898 (1997).

5. L.F. Cugliandolo and J. Kurchan, Physica A {\bf 263}, 242 (1999).

6. T.S. Grigera and N.E. Israeloff, Phys. Rev. Lett. {\bf 83}, 5038 (1999).

7. D. H\'erisson and M. Ocio, Phys. Rev. Lett. {\bf 88}, 257202 (2002); preprint cond-mat/0403112.

8. L. Buisson, M. Ciccotti, L. Bellon and S. Ciliberto, preprint cond-mat/0403294.

9. B. Abou and F. Gallet, preprint cond-mat/0403561.

10. N. Pottier, Physica A {\bf 317}, 371 (2003).

11. N. Pottier and A. Mauger, Physica A {\bf 332}, 15 (2004).

12. T.G. Mason and  D.A. Weitz, Phys. Rev. Lett. {\bf 74}, 1250 (1995).

\bye